Building Communication Skills in a Theoretical Statistics Course

Amy Wagaman

Department of Mathematics and Statistics, Amherst College, P.O. Box 5000, Amherst, MA 01002

Abstract

The "traditional" theoretical statistics course which develops the theoretical underpinnings of the discipline (usually following a probability course) is undergoing near-continuous revision in the statistics community. In particular, recent versions of this course have incorporated more and more computation. We take a look at a different aspect of the revision - building student communication skills in the course, in both written and verbal forms, to allow students to demonstrate their ability to explain statistical concepts. Two separate projects are discussed, both of which were engaged in by a class of size 17 in Spring 2015. The first project had a computational aspect (performed using R), a statistical theory component, and a writing component, and was based on the historical German tank problem. The second project involved a class presentation and written report summarizing, critiquing, and/or explaining an article selected from The American Statistician.

Introduction

In our statistics courses, our students learn quite a bit about both data analysis and the theory behind statistics. They learn to create visualizations, perform hypothesis testing and generate confidence intervals, discuss issues related to experimental design, derive maximum likelihood estimators and posterior distributions, and much more. With the advances in technology and statistical software, the amount of computation in our courses has increased, with students performing simulations (often using software, such as R (R 2009) to verify results or perform randomization-based procedures. These skills in statistics and in computing will serve our students well, but there is another skill set that we need to help our students acquire.

Our statistics students need to practice and build strong communication skills. Being able to perform an analysis or suggest a reasonable model means little if the student cannot communicate their results and their reasoning to others, including audiences with less of a statistics vocabulary. The importance of communication skills for statistics students is highlighted in the recent Curriculum Guidelines for Undergraduate Programs in Statistical Science which suggests programs provide opportunities for students to practice communication skills, and learn about ethics (Guidelines 2014). In the spirit of the guidelines, we include communication skills as a learning outcome for our statistics majors, and have worked to be sure our program provides multiple avenues for students to practice these skills.

Our students engage in a variety of activities to practice communication skills including presentations (both formal and informal, with slides and speaking components), handout creation and use, and report writing (with varying length). Skills developed include writing ability (for reports, slides, and handouts), speaking ability (maintaining eye contact, voice volume, etc.), and the ability to create good visualizations (for inclusion in reports, handouts, and slides).

In this article, we explore how two course projects in a theoretical statistics course allowed students to build and practice communication skills. We provide a brief background about the course in Section 1. We describe the two course projects in Sections 2 and 3, respectively. Finally, we conclude with some discussion in Section 4.

## 1. Course Background

The theoretical statistics course discussed here is a 300-level (previously listed as a 400-level) course with a primary audience of juniors and seniors. Statistics majors take the course as juniors in preparation for a senior capstone experience in statistics. Students from other majors, including majors in mathematics and economics, take the course as well. The course has a pre-requisite of one semester of probability. Topics covered in the course include: Bayesian inference, maximum likelihood estimation, sufficient statistics, confidence intervals, hypothesis testing and test selection, non-parametric procedures, and linear models. The course uses the statistical computing software R (with RStudio and RMarkdown) for computations, simulations, analysis, etc. (R 2009).

## 2. Course Project I - Tanks, Anyone?

The first course project covers the topics of estimation and simulation. It is based on the historical "German tank problem" and the associated common estimation exercise used to engage students in this course (multiple examples of related projects may be found online).

### 2.1 Project Setup and Student Directions

Students are presented with the project instructions and given about three weeks to work on their submissions. A handout describes the project. In short, students are told that they are interested in the following problem: A random sample of $k$ values from a population with individuals labeled from 1 to $N$ is drawn. An estimate for $N$ is needed. Students are also provided with a small data set of $k$ values for which the estimate of $N$ is needed.

Students are directed to derive several estimators (and examine their properties - expectation and variance) for $N$, including the method of moments estimator and the maximum likelihood estimator. Some hints are provided about trickier derivation pieces. Students are also instructed to brainstorm additional estimators, and use simulation (in R) to compare all their estimators. Due to one student's weak R background, the class was provided with example simulation code for one nonsensical estimator.

The students are then tasked with writing a "report" explaining their choice of "best" estimator with support, via their calculations and the simulations. The report is actually framed as a letter to their commanding officer (as in an intelligence officer), which the students seem to enjoy the creative aspect of. Next, we explore the various communication aspects involved in the project.

2.2 Communication Practice, Outcomes, and Feedback

There are a variety of communication skills that students need to develop and practice. In this project, the students are focusing on writing, rather than speaking, skills. The writing includes a letter, with a summary of findings, appropriate supporting information, including derivations and simulation results, as well as the simulations where the students write code (though students did not submit their code for this, due to being given example code).

Generally, students did not seem to have any issues with creating the letter, writing about their findings, and comparing the estimators via their derivations. (For example, it seems very simple for students to write sentences like: the variance of estimator one is less than that of estimator two, so I prefer estimator one). There were, of course, minor issues across the board with spelling and proofreading, along with some organizational issues. However, students had more issues when writing about their simulations, and using appropriate support for their chosen estimator in their reports. This issue was discussed via an eCoTs virtual poster (Wagaman 2016), but more detail follows here.

To understand the problems encountered, we consider what students should be able to do in regards to their simulations. First, students should be able to explain what their simulation does. They should be able to run it and obtain results, helping them to choose a preferred estimator. Then, students should be able to extract meaningful support for their estimator from their simulation. Once students have identified what support to include, that support should be provided appropriately in the written document. Finally, in this particular case, since example code was provided (with example settings of $N$=300 and $k$=15), students should convey (in some form) how they explored different settings for the simulation other than those provided. In other words, their chosen support should include settings other than $N$=300 and $k$=15.

It should come as no surprise that most students decided to summarize their simulation results with tables of descriptive statistics, supplemented with selected graphs. This is in fact, how many of us would choose to display our findings. However, students had issues with writing about these results. In particular, some students found writing about their tables very challenging. On one extreme, a student included a table and simply stated that the table was support for their estimator. At the other extreme, students were providing multiple sentences about their tables (at least one sentence per table row), including the majority of values from the table, making the text completely redundant. Based on this anecdotal experience, it is clear that students may need some instruction about writing about tables. For example, when preparing a table for a paper and writing the accompanying text, students may need to think about: "What is too obvious to restate?" and "What is useful to point out to the reader?".

In terms of graphs, it appears that some students need more guidance trying to prepare figures for reports. For example, some students included too many graphs (in my opinion), and in inappropriate formats (e.g. 16 graphs on a page, too tiny to really examine). Others included an appropriate number of graphs but had organizational issues (e.g. 1 graph per page for 6 pages). Appropriate labels and titles were problems for some students.

After projects were submitted, they were assessed. Students received a copy of the assessment rubric with notes about their work when projects were returned. Organizational and formatting comments were provided to assist students with the problem areas described above, but students were not allowed to submit a revision of this project.

## 3. Course Project II - Working with TAS Articles

Students need to be able to apply their knowledge to new situations, and our programs should provide some practice with this task. In particular, students should be able to apply their knowledge when reading new literature in the field they are studying. This course project was designed to help students read a new article (at an appropriate level), process the material, and demonstrate their understanding via a class presentation and short written report. Students were allowed to revise their submission, with access to the assessment rubric.

### 3.1 Project Setup and Student Directions

For this second course project, students choose an article from The American Statistician (TAS) from a curated list supplied by me (with articles appropriate based on their background from the course). The students then read and worked through their selected article. In the event they encountered an unfamiliar term or method, they were to do research to be able to explain it to their classmates. After processing their article, students were tasked with presenting their article to the class (a six minute presentation) and writing a 4-6 page (double spaced) summary of what they learned, what methods were used, etc. In both these assignments, the students might have to pick and choose what aspects of the article they shared with the class, due to length of their articles and need to explain new concepts to their classmates. Students were encouraged to come up with examples to demonstrate methods from their articles to the class. Simulations could also be used to verify results from the articles (or generate examples), and some students sought assistance with creating simulations in R.

### 3.2 Communication Practice, Outcomes, and Feedback

Students were working on both a presentation and a written report for this project. For the presentation, most students made slides or prepared a handout, and selecting what to display to the audience was challenging for some. For both the presentation and the written report, one key issue encountered by all students was introducing their audience to a new setting (whatever setting was described in their article).

The student presentations occurred a few days before the written report was due. Indeed, student presentations were spread over three class days (themed based on the articles chosen). Students provided feedback to one another via a comment sheet (collected by the instructor, and results distributed to speakers anonymously). The comment sheet included the following questions:

- How well did the presenter convey the statistical topic at hand? Were new concepts clearly explained?
- How well did the presenter maintain your interest? Were you engaged in the presentation?
- Overall rating of presentation

Students were also able to receive other comments (space was provided at the bottom of the page) from their classmates. As the instructor, I filled out a similar sheet paying attention to the statistical concepts conveyed and the presentation components (eye contact, organization, etc.). Students were able to incorporate their presentation feedback into their written reports. This was very important for some students to realize that additional explanation of the setting and terms might be useful for a reader.

Reports were submitted, assessed, and returned to the students fairly promptly. The assessment rubric for the report included the following areas (which were allocated different point values):

- General: Writing, spelling, proofreading, grammar, etc.
- Citations (at a minimum, the article itself must be cited).
- Topic: Are main topics from the article conveyed? Is an appropriate subset chosen for discussion if the article was very long?
- Explanation: How well are new concepts explained? Was appropriate research/background information obtained and conveyed? Is the setting introduced appropriately? Are terms/notation defined before they are used?
- Audience: Is the writing appropriate for the target audience (another student in the course)?
- Statistics: Are there any issues in the presentation of the statistical topics/issues?
- Interest/Creativity/Effort/Visuals: For example, is the interest level of the reader maintained? Were simulations used to illustrate results? Were appropriate visuals chosen?

Students received a copy of this rubric with their assessment before revisions, so they could clearly see what areas they had struggled in. Students were then allowed to submit a revised report to address comments, and could earn some points back (up to half of what was originally lost). Revisions were undertaken by 10 of the 17 students. Just as a note, reports were requested to be between 4 and 6 pages long. This meant that many students had to pick and choose what aspect of the selected paper they wanted to write about in their reports. In the end, a few students went over the page limit, though this was not a major concern for me for this assignment. But it did make me think of other assignments where enforcing a page limit might make sense and be useful assignments for the students to engage in.

Anecdotally, most students lost points in either the General, Explanation, or Statistics assessment areas. The challenge of explaining a new setting was particularly hard for some students. However, these issues are addressable in revisions, and all students undertaking a revision were able to successfully address at least some of the original concerns with their submissions.

3.3 Materials

For relevant materials for this project, including the list of selected articles from Spring 2015, and example rubrics, please contact the author.

4. Discussion

The projects as described were completed in a class of 17 students. A number of easy adaptations could make these projects more deployable for your classrooms. In both cases, the projects can easily be made to be group assignments instead of individual assignments. The instructor could randomly assign groups or insist groups change between projects.

For the first course project on the German tank problem, the project could be adapted with more responsibility on students to perform the simulation. Indeed, example code need not be supplied. Instead, the instructor could require students to write and submit their own simulation code. For the writing aspect, a revision could be incorporated to allow students to improve their submissions. Finally, based on the issues students included, instructors could consider incorporating an activity in the course prior to this project about writing about simulations or tables to help students prepare for that component.

For the second course project on presenting and writing about a selected article, a number of adaptations are possible based on tailoring the article list. For example, the article list be focused on specific topics such as Bayesian inference or confidence intervals. Simulations could be required which would impact the selection of articles for the provided list to students. Similar activities could be done using different journals/magazines depending on the level of the audience (e.g. Significance).

In summary, both course projects involved communication aspects ranging from written reports, writing about simulations, and class presentations (handout and slide creation, presentation skills) that allowed students to practice their communication skills as well as deepen their statistical understanding.

This course is undergoing revision at many institutions. For more thoughts on revisions to mathematical statistics courses, the reader is encouraged to see Green and Blankenship (2015).